\documentclass[aps,pre,preprint,showpacs]{revtex4}
\usepackage[dvips]{graphicx}
\usepackage{amsmath}
\usepackage{amsfonts}
\usepackage{amssymb}
\usepackage{latexsym}
\usepackage{epsfig}
\usepackage{oldgerm}
\usepackage{theorem}
\newcommand{\mib}[1]{\mbox{\boldmath $#1$}}

\def\C{\mathbb{C}}
\def\R{\mathbb{R}}
\def\H{\mathbb{H}}
\def\D{\mathbb{D}}
\def\Z{\mathbb{Z}}
\def\vsigma{\mib{\sigma}}

\begin{document}

\title{Iterative Schwarz-Christoffel Transformations \\
Driven by Random Walks and Fractal Curves}

\author{Fumihito Sato}
\affiliation{Department of Physics,
Faculty of Science and Engineering,
Chuo University, Kasuga, Bunkyo-ku, Tokyo 112-8551, Japan}
\author{Makoto Katori} 
\email[]{katori@phys.chuo-u.ac.jp}
\affiliation{Department of Physics,
Faculty of Science and Engineering,
Chuo University, Kasuga, Bunkyo-ku, Tokyo 112-8551, Japan}

\date{31 March 2010}

\begin{abstract}
Stochastic Loewner evolution (SLE) is a differential equation
driven by a one-dimensional Brownian motion (BM),
whose solution gives a stochastic process of conformal
transformation on the upper half complex-plane $\H$.
As an evolutionary boundary of image of the transformation,
a random curve (the SLE curve) is generated,
which is starting from the origin and running in $\H$
toward the infinity as time is going.
The SLE curves provides a variety of statistical ensembles
of important fractal curves, if we change the
diffusion constant of the driving BM.
In the present paper, we consider 
the Schwarz-Christoffel transformation (SCT),
which is a conformal map from $\H$ to 
the region $\H$ with a slit
starting from the origin.
We prepare a binomial system of SCTs,
one of which generates a slit in $\H$
with an angle $\alpha \pi$ from the positive direction
of the real axis,
and the other of which with an angle $(1-\alpha) \pi$.
One parameter $\kappa >0$ is introduced to
control the value of $\alpha$ and the length of slit.
Driven by a one-dimensional random walk, which is
a binomial stochastic process,
a random iteration of SCTs is performed.
By interpolating tips of slits
by straight lines, we have a random path in $\H$,
which we call an Iterative SCT (ISCT) path.
It is well-known that,
as the number of steps $N$ of random walk
goes infinity, each path of random walk divided by $\sqrt{N}$
converges to a Brownian curve.
Then we expect that the ISCT paths divided by $\sqrt{N}$
(the rescaled ISCT paths)
converge to the SLE curves in $N \to \infty$.
Our numerical study implies that, for sufficiently large $N$,
the rescaled ISCT paths will have the same statistical properties
as the SLE curves have, supporting our expectation.
\end{abstract}


\maketitle

\section{INTRODUCTION}

One of the highest topics of recent progress in statistical
physics of critical phenomena and random fractal patterns
is introduction of the Stochastic Loewner Evolution (SLE)
by Schramm \cite{Sch00,Law05,Law07}.
The SLE will provide a unified theory of statistics
of random curves in the plane,
which covers (continuum limits of)
random interfaces characterizing surface
critical phenomena in equilibrium
({\it e.g.} the percolation exploration process,
the critical Ising interface),
models of random chains in polymer physics
({\it e.g.} the self-avoiding walks),
fractal patterns playing important roles in
non-equilibrium statistical mechanics models
({\it e.g.} the loop-erased random walks 
and the uniform spanning trees for sandpile models
and forest fire models showing self-organized criticality), 
and so on \cite{KN04,Car05,BB06}.
The theory is based on two branches of 
mathematics, the complex function theory \cite{Pom75}
and the stochastic analysis \cite{Law05,Law07}
and is strongly connected with the conformal field theory
\cite{Car01,BB03,FW03}. 

As well as by wideness of applications 
and by richness of mathematics,
we are attracted by simple setting
of the theory ;
Consider a complex plane $\C$.
(i) First we consider a motion of Brownian particle
on the real axis $\R$. We assume that it starts
from the origin 0 at time $t=0$ and the diffusion 
constant is given by $\kappa > 0$.
If we write the position of the diffusion particle
on $\R$ at time $t \geq 0$ as
$U_t$, then 
$\langle U_t \rangle=0$ and
$\langle U_t^2 \rangle= \kappa t$
for $t \geq 0$.
Usually we denote the position of a one-dimensional
standard Brownian motion (BM) at time $t$
by $B_t$, for which
$\langle B_t \rangle \equiv B_0=0$,
$\langle B_t^2 \rangle =t, t \geq 0$.
The BM has the scaling property such that
for any constant $c > 0$, the distribution
of the position of BM at time $c^2 t$
is equal to that of the position of
BM at time $t$ multiplied by a factor $c$, that is, 
the equality $B_{c^2 t}=c B_t$ holds 
in distribution.
Then we can give the above $U_t$ by
\begin{equation}
U_t=\sqrt{\kappa} B_t, \quad t \geq 0.
\label{eqn:Ut1}
\end{equation}
(ii) Then we solve the following partial differential
equation for a complex function $f_t(z)$ 
on the upper half complex-plane
$\H=\{z \in \C : {\rm Im} \, z > 0\}$,
\begin{equation}
\frac{\partial f_t(z)}{\partial t}
=- \frac{\partial f_t(z)}{\partial z}
\frac{2}{z-U_t}
\label{eqn:SLE1}
\end{equation}
under the initial condition
$f_0(z)=z$.
(iii) Note that the boundary of $\H$
consists of the real axis $\R$ and an infinity point.
When $z \in \H$ approaches the special point $U_t$ on $\R$,
the position of the BM,
the RHS of (\ref{eqn:SLE1}) diverges.
If we trace the image of this singular point
\begin{equation}
\gamma_t=f_t(U_t),
\quad t \geq 0
\label{eqn:SLE2}
\end{equation}
we will have a curve 
$\gamma(0,t] \equiv \{\gamma_s : 0 < s \leq t\}$
in $\H$ starting from the origin.
For each $s >0$ 
the curve $\gamma(0,s]$
and the region enclosed by parts of the curve and
the real axis $\R$ should be eliminated from $\H$
in order to continue to solve Eq.(\ref{eqn:SLE1})
for $t > s$.

For any deterministic simple curve
$\gamma(0,t]$ in $\H, t \geq 0$,
Loewner proved that there exists a real-valued
function $U_t$ and a conformal map $f_t$,
which is one-to-one from $\H$
to $\H \setminus \gamma(0,t]
\equiv \{z \in \H: z \not= \gamma_s, 0 < s \leq t\}$, 
the upper half complex-plane with a slit $\gamma(0,t]$,
such that $f_t$ and $U_t$ solve Eq.(\ref{eqn:SLE1})
with the condition (\ref{eqn:SLE2}).
The equation (\ref{eqn:SLE1}) is called
the Loewner equation \cite{Pom75,SLEa}. 
Schramm considered an inverse problem:
given $U_t$ on $\R$ and derive
a curve $\gamma(0,t]$ by solving Eq.(\ref{eqn:SLE1}).
Since he gave $U_t$ by a BM as Eq.(\ref{eqn:Ut1}),
$\gamma_t$ given by Eq.(\ref{eqn:SLE2}) 
performs a stochastic motion and 
the obtained curve $\gamma(0,t]$ is statistically distributed 
in $\H$ \cite{Sch00}. Equation (\ref{eqn:SLE1})
driven by a BM with variance $\kappa t$
is called the stochastic Loewner equation
or the Schramm-Loewner evolution (SLE) \cite{SLEb},
and a random curve $\gamma(0,t], t \geq 0$
is called the SLE curve with parameter $\kappa$
(the SLE$_{\kappa}$ curve).

Although the change of diffusion constant $\kappa$ causes
only quantitative change of the driving function
$U_t$, that is, scale/time change
$\sqrt{\kappa} B_t=B_{\kappa t}$ in distribution,
it does qualitative change of SLE$_{\kappa}$ curves.
When $0 < \kappa \leq 4$, the curve is simple
(with no self-intersection) 
with $\gamma(0, \infty) \subset \H$.
When $4 < \kappa < 8$, the curve is self-intersecting
and $\gamma(0,\infty)$ hits 
the real axis $\R$ infinitely many times, 
but it is not dense on $\H$ :
$\gamma(0, \infty) \cap \H \not= \H$
with $\lim_{t \to \infty} |\gamma_t|=\infty$.
And when $\kappa \geq 8$,
it will cover whole of $\H$ in $t \to \infty$
\cite{Law05,Law07}.
The fractal dimension (Hausdorff dimension)
of the SLE$_{\kappa}$ curve is determined 
as \cite{Bef08,Law09}
\begin{equation}
d^{\kappa}=\left\{
\begin{array}{ll}
1+\kappa/8, & \quad 0 < \kappa < 8 \cr
2, & \quad \kappa \geq 8.
\end{array} \right.
\label{eqn:d1}
\end{equation}
Moreover, if $\kappa$ is chosen to be a special value,
SLE$_{\kappa}$ curves provide the statistical ensembles
of continuous limits of random discrete paths
studied in statistical mechanics models 
exhibiting critical phenomena
or in fractal models on lattices. 
For example, the value $\kappa=6$ is for the
critical percolation model
\cite{Smir01,Bef07}.
Effective methods for numerical simulations of 
SLE$_{\kappa}$ curves by computers 
have been reported \cite{Ken07,Ken09,KMN09}.

Today we can learn from mathematics literatures
that the Loewner equation (\ref{eqn:SLE1}) for 
a deterministic function $U_t$ had played important
roles in the complex function theory even before
Schramm introduced its stochastic version \cite{Pom75}.
We shall say, however, that
this equation has not been familiar to us,
physicists.
More familiar differential equation to us in the
complex calculus is the one,
whose solution gives 
a conformal transformation from $\H$ to 
the interior of a polygon on the complex plane
with mapping the real axis $\R$ 
to a piecewise linear boundary of the polygon,
called the Schwarz-Christoffel transformation (SCT)
(see, for example, \cite{AF03}).
So here we try to discuss the Loewner equation (\ref{eqn:SLE1})
by using a special case of SCT.

\begin{figure}[htbp]
  \begin{center}
   \includegraphics[width=100mm]{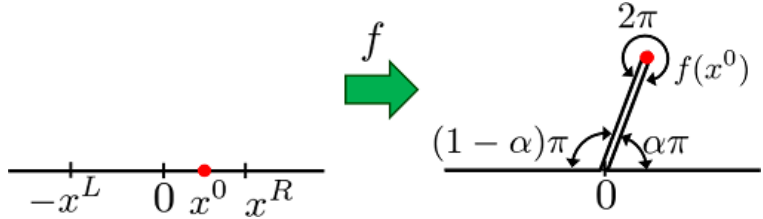}
  \end{center}
  \caption{SCT generating a straight slit in $\H$}
  \label{fig:fig_SCT}
\end{figure}
Let $0 < \alpha < 1$.
Consider a conformal map $f$ from $\H$
to the upper half complex-plane with a straight slit
starting from the origin, 
$\H \setminus \{\mbox{a slit}\}$, where the angle 
between the slit and the positive direction of 
the real axis is $\alpha \pi$ 
as shown in Fig.\ref{fig:fig_SCT}.
Since $\H$ with the straight slit can be regarded
as a polygon with the interior angles
$(1-\alpha) \pi$ on the left side of the origin,
$2 \pi$ around the tip of the slit,
and $\alpha \pi$ on the right side of the origin, 
for any length of a slit, the conformal map
is given as an SCT, which solves 
the differential equation
\begin{equation}
\frac{df(z)}{dz}
=b_1 (z+x^{\rm L})^{-\alpha}
(z-x^0)(z-x^{\rm R})^{\alpha-1},
\label{eqn:SC1}
\end{equation}
where $b_1$ is a complex number
and $x^{\rm L}, x^{0}, x^{\rm R}$ are real numbers
satisfying the inequalities,
$x^{\rm L}>0, x^{\rm R}>0,
-x^{\rm L} < x^{0} < x^{\rm R}$.
By this transformation, both of the points
$-x^{\rm L}$ and $x^{\rm R}$ on $\R$
are mapped to the origin,
$f(-x^{\rm L})=f(x^{\rm R})=0$,
and $x^{0}$ to the tip of the slit.
See Fig.\ref{fig:fig_SCT}.
We have found the general solution of (\ref{eqn:SC1}) 
expressed by
\begin{eqnarray}
&& f(z)=b_0+b_1 (z+x^{\rm L})^{1-\alpha}
(z-x^{\rm R})^{\alpha}
\nonumber\\
&& \quad \times
\left\{ 1+
\frac{x^{0}-x^{\rm R}+\alpha(x^{\rm L}+x^{\rm R})}{x^{\rm L}+x^{\rm R}}
\frac{\Gamma(1-\alpha)}{\Gamma(2-\alpha)}
F\left(1,1, 2-\alpha;
\frac{x^{\rm L}+z}{x^{\rm L}+x^{\rm R}} \right) \right\},
\label{eqn:SC2}
\end{eqnarray}
where $b_0$ is a complex number,
$\Gamma(z)=\int_0^{\infty} e^{-u} u^{z-1} du$
(the gamma function),
and $F(\alpha, \beta, \gamma;z)$ is 
Gauss's hypergeometric function
$F(\alpha, \beta, \gamma;z)
=\Gamma(\gamma)/\{\Gamma(\alpha) \Gamma(\beta)\}
\sum_{n=0}^{\infty} 
\Gamma(\alpha+n) \Gamma(\beta+n)
z^{n}/\{\Gamma(\gamma+n) n! \}$.
We impose the hydrodynamic condition
\begin{equation}
\frac{f(z)}{z} \to 1
\quad \mbox{in} \quad z \to \infty.
\label{eqn:hydro1}
\end{equation}
Then $b_0=0, b_1=1$ and
$x^0-x^{\rm R}+\alpha(x^{\rm L}+x^{\rm R})=0$, 
and we have $f(z)=(z+x^{\rm L})^{1-\alpha}
(z-x^{\rm R})^{\alpha}$.
Eq.(\ref{eqn:SC1}) is rewritten in this case as
\begin{equation}
\frac{df(z)}{dz} \frac{2}{z-x^{0}}
=\frac{2 f(z)}{(z+x^{\rm L})(z-x^{\rm R})}.
\label{eqn:SC3}
\end{equation}
We then introduce a parameter ``time" $t \geq 0$ and assume that
$x^{\rm L}$ and $x^{\rm R}$, and thus also $x^0$, 
depend on $t$ by setting
\begin{equation}
f_t(z)=(z+x^{\rm L}_t)^{1-\alpha}
(z-x^{\rm R}_t)^{\alpha}.
\label{eqn:SC4}
\end{equation}
The differential of $f_t$ with respect to $t$ is written as
$$
\frac{\partial f_t(z)}{\partial t}
=-\frac{2 A(z,t) f_t(z)}
{(z+x^{\rm L}_t)(z-x^{\rm R}_t)}
$$
with
$A(z,t)=[-(1-\alpha)(z-x^{\rm R}_t) dx^{\rm L}_t/dt
+\alpha (z+x^{\rm L}_t)dx^{\rm R}_t/dt]/2$.
Let $x^{\rm L}_t=2 c t^{\beta}, x^{\rm R}_t=2t^{\beta}/c$
with constants $c>0$ and $\beta$.
Then we can see that, if and only if
$c=\sqrt{\alpha/(1-\alpha)}$ and $\beta=1/2$,
$A(z,t)$ becomes independent both of $z$ and $t$;
$A(z,t) \equiv 1$.
Combining this observation with Eq.(\ref{eqn:SC3})
gives the following result:
For $0 < \alpha < 1$, the SCT
\begin{equation}
f^{\alpha}_t(z)=\left(z+ 2 \sqrt{\frac{\alpha}{1-\alpha}}
\sqrt{t} \right)^{1-\alpha}
\left(z- 2 \sqrt{\frac{1-\alpha}{\alpha}}
\sqrt{t} \right)^{\alpha}
\label{eqn:SC5}
\end{equation}
is not only a solution of the Schwarz-Christoffel
differential equation (\ref{eqn:SC1}),
but also of the Loewner equation (\ref{eqn:SLE1})
with the driving function
\begin{eqnarray}
U_t^{\alpha} &=& x^0_{t}
=x^{\rm R}_{t}-\alpha(x^{\rm L}_{t}+x^{\rm R}_{t})
\nonumber\\
&=& \left\{ \begin{array}{ll}
\sqrt{\kappa^{\alpha} t} & 
\quad \mbox{if $\alpha \leq 1/2$} \cr
-\sqrt{\kappa^{\alpha} t} & 
\quad \mbox{if $\alpha > 1/2$},
\end{array} \right.
\label{eqn:SCU}
\end{eqnarray}
where
\begin{equation}
\kappa^{\alpha}=\frac{4(1-2 \alpha)^2}{\alpha (1-\alpha)}.
\label{eqn:SCK}
\end{equation}
As time $t$ goes, the straight slit performs as
an ``evolutionary boundary" of the image of $\H$ by $f_t$,
in which the tip of the slit (\ref{eqn:SLE2})
is evolving as
\begin{eqnarray}
\gamma_t^{\alpha} &=& f^{\alpha}_t(U_t^{\alpha})
\nonumber\\
&=& 2 \left( \frac{1-\alpha}{\alpha} \right)^{1/2-\alpha}
e^{i \alpha \pi} \sqrt{t},
\quad t \geq 0.
\label{eqn:SCgamma}
\end{eqnarray}

One can observe the equality
\begin{equation}
U_t^{\alpha}= \pm \sqrt{
\langle ( \sqrt{\kappa^{\alpha}} B_t)^2 \rangle},
\quad t \geq 0,
\label{eqn:SCU2}
\end{equation}
since $\langle B_t^2 \rangle = t$.
That is, the driving function $U_t^{\alpha}$ for the
above SCT (\ref{eqn:SCU}) is the positive or the negative 
root square of the squared average of the
random driving function (\ref{eqn:Ut1}) of the SLE.
So if we are able to introduce fluctuations into the SCT
systematically, we can draw approximations of SLE curves
on $\H$.
It may be the basic idea to simulate SLE curves
by dividing a time period $(0,t]$ into
$n$ small intervals $\{(t_{j-1}, t_{j}] : j=1,2,\dots, n\}$
by setting $0=t_0 < t_1 < \cdots < t_n=t$
and the above single SCT is replaced by an $n$-multiplicative
map of infinitesimal SCTs with sufficiently large $n$
\cite{Ken07,Ken09}.

\begin{figure}[htbp]
  \begin{center}
   \includegraphics[width=100mm]{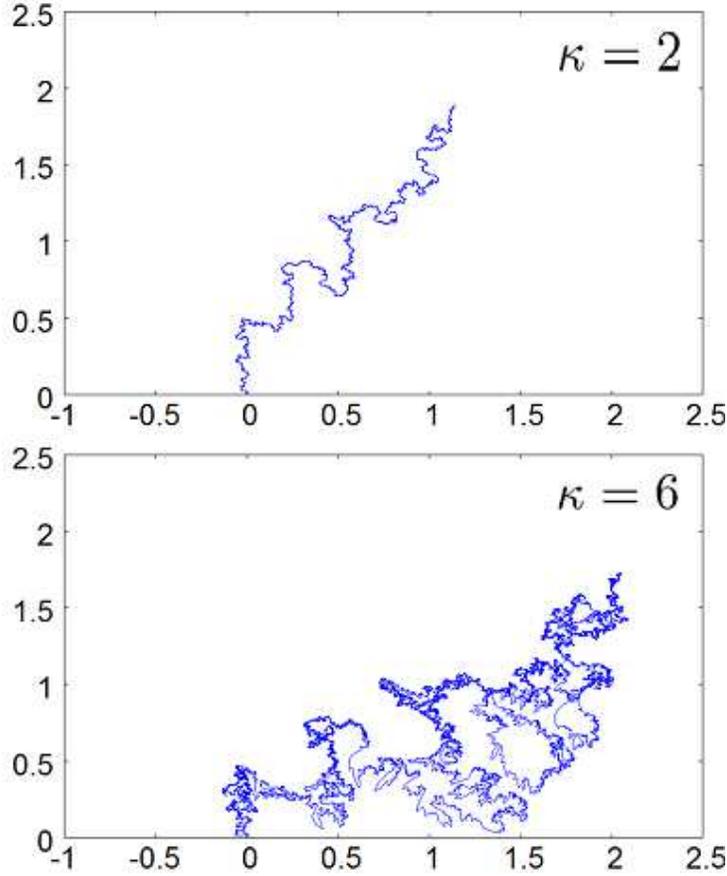}
  \end{center}
  \caption
  {Approximations of the SLE$_{\kappa}$ curves up to time $T=1$
  expressed by the ISCT$_{\kappa}$ paths
  for $\kappa=2$ and $\kappa=6$.}
  \label{fig:fig_path50K}
\end{figure}

On the other hand, we know the fact that
the diffusion property of BM can be
observed in long-time asymptotic behavior
of a simple discrete-time stochastic process,
random walk (RW).
Let $\sigma_j, j=1,2,3, \dots$ be independent and
identically distributed (i.i.d.) random variables
taking values 
$\sigma_j=1$ with probability 1/2
and $\sigma_j=-1$ with probability 1/2.
Consider a simple symmetric RW
on the one-dimensional lattice 
$\Z=\{ \dots, -2,-1,0,1,2, \dots\}$
starting from the origin 0 at time $n=0$.
We denote the position of the random walker 
at time $n=0,1,2, \dots$
by $w_n$. Then, $w_0=0$ and
\begin{equation}
w_n=\sum_{j=1}^{n} \sigma_j, \quad n=1,2,3, \dots.
\label{eqn:Wt}
\end{equation}
In the present paper, we consider an SCT as a functional of
a random variable $\sigma$ and consider an
Iterative system of SCTs (ISCTs) driven by RW.
By this system, each time series of steps 
$(\sigma_1, \sigma_2, \dots )$ of RW is mapped to
a series of points $(\xi_1, \xi_2, \dots)$ in $\H$.
Let $W_n$ and $\Xi_n, n \geq 0$ be the
interpolations by straight lines 
of $w_n$ and $\xi_n, n=0,1,2, \dots$, respectively.
Since for $T > 0$,
$\{W_{Nt}/\sqrt{N} : 0 \leq t \leq T \}$
converges to $\{B_t : 0 \leq t \leq T \}$ as
$N \to \infty$ in distribution, 
\begin{equation}
\zeta^N(0,T]=\left\{
\frac{\Xi_{Nt}}{\sqrt{N}}, 0 < t \leq T 
\right\}, \quad T > 0,
\label{eqn:zetaN}
\end{equation}
which we call the rescaled ISCT path with $N$
up to time $T$, 
will converge to an SLE curve up to time $T$,
$\gamma(0,T]$, in distribution as $N \to \infty$.
Figure \ref{fig:fig_path50K}
shows the rescaled ISCT paths
$\zeta^N(0,1]=\{\Xi_{Nt}/\sqrt{N}, 0 \leq t \leq 1 \}$
with $N= 5 \times 10^{4}$ for $\kappa=2$ and 6, 
which are drawn by interpolating 
the series of points $\{\xi_j/\sqrt{N}\}_{j=0}^{N}$
by lines.
They seems to approximate the SLE$_{\kappa}$ curves
$\gamma(0,1]$ very well.
In the present paper, we will show that, 
for sufficiently large $N$,
$\zeta^N_t$ has the same statistical 
properties as the SLE curve $\gamma_{t}$ has.

The paper is organized as follows.
In Sec.II, we define the ISCT driven by random walk 
with a parameter 
$\kappa=\kappa^{\alpha}$ defined by (\ref{eqn:SCK})
and the ISCT$_{\kappa}$ paths.
In Sec.III, by observing the behavior of ISCT paths
$\Xi(0,n]$ for small values of $n$,
we show that, even if we use the same realization
of RW as a driving function,
the ISCT$_{\kappa}$ paths with large values of $\kappa$ 
exhibit much more complicated motion on $\H$
than those with smaller values of $\kappa$.
In Sec.IV, we report the properties
of the rescaled ISCT paths $\zeta^{N}(0,T]$
with $T=1$ based on large scaled computer
simulations with $N \simeq 10^4 \sim 10^{5}$.
We evaluate the fractal dimensions $d^{\kappa}_{\rm ISCT}$
of the $N \to \infty$ limits of
$\zeta^{N}(0,1]$ for several values of 
$\kappa$, 
and the dependence of $d^{\kappa}_{\rm ISCT}$
on $\kappa$ is compared with that of the fractal dimensions
$d^{\kappa}$ of the SLE curves given by (\ref{eqn:d1}).
We also show that, for $4 < \kappa < 8$,
the generalized version of Cardy's formula
of SLE$_{\kappa}$ curves 
\cite{Car92,Law05} will be applicable to 
the rescaled ISCT$_{\kappa}$ paths 
$\zeta^{N}(0,T]$, if
$N$ and $T$ are sufficiently large.
Section V is devoted to give concluding remarks.
Appendix A is prepared for giving recurrence relations,
which will be useful to analyze the ISCT paths.

\section{ISCT driven by RW}

Noting that Eq.(\ref{eqn:SCK}) is solved for $\alpha$ as
$\alpha=[1\pm \sqrt{\kappa/(\kappa+16)}]/2$, we set
\begin{equation}
\alpha^{\kappa}(\sigma)=\frac{1}{2}
\left[ 1- \sigma \sqrt{\frac{\kappa}{\kappa+16}} \right]
\label{eqn:alpha1}
\end{equation}
for $\kappa > 0, \sigma \in \{-1,1\}$,
and define the SCT as a functional of a random 
variable $\sigma$ by
\begin{eqnarray}
F^{\kappa}_{\sigma}(z)
&\equiv& f^{\alpha^{\kappa}(\sigma)}_{1}(z)
\nonumber\\
&=& \left( z + 2 \sqrt{\frac{\alpha^{\kappa}(\sigma)}
{1-\alpha^{\kappa}(\sigma)}} \right)^{1-\alpha^{\kappa}(\sigma)}
\left( z-2 \sqrt{\frac{1-\alpha^{\kappa}(\sigma)}
{\alpha^{\kappa}(\sigma)}} \right)^{\alpha^{\kappa}(\sigma)}.
\label{eqn:F1}
\end{eqnarray}

\begin{figure}[htbp]
  \begin{center}
   \includegraphics[width=100mm]{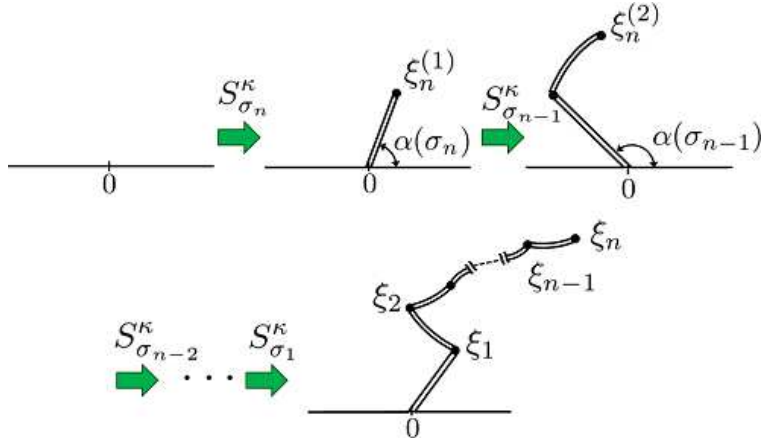}
  \end{center}
  \caption{Iteration of SCTs}
  \label{fig:fig_ISCT}
\end{figure}

Given one step $\sigma_1$ of RW on $\Z$,
we consider an SCT, $F^{\kappa}_{\sigma_1}(z)$,
which is a conformal map from $\H$ to
the region $\H$ with a straight slit.
The straight slit starts from the origin and ends at
the tip located at
\begin{eqnarray}
\xi_1 &=& F^{\kappa}_{\sigma_1}
(\sqrt{\kappa} \sigma_1) \nonumber\\
&=& 2 \left( \frac{1-\alpha^{\kappa}(\sigma_1)}
{\alpha^{\kappa}(\sigma_1)} \right)^{1/2-\alpha^{\kappa}(\sigma_1)}
e^{i \alpha^{\kappa}(\sigma_1) \pi}.
\label{eqn:gamma1}
\end{eqnarray}
Next assume that two steps of RW, $(\sigma_1, \sigma_2)$,
is given.
We transform $\H$ by an SCT, $F^{\kappa}_{\sigma_2}$.
The image of $\H$, $F^{\kappa}_{\sigma_2}(\H)$, is
the region $\H$ with the straight slit, 
which starts from the origin
and ends at 
$\xi_2^{(1)} \equiv F^{\kappa}_{\sigma_2}(\sqrt{\kappa} \sigma_2)$.
Then we consider the transformation of the region
$\H$ with this straight slit by another SCT, $F^{\kappa}_{\sigma_1}$.
By this SCT, a straight slit from the origin to the point $\xi_1$
is generated as shown by (\ref{eqn:gamma1}).
The image by $F^{\kappa}_{\sigma_1}$ 
of the straight slit between the origin
and $\xi_2^{(1)}$ in $\H$ is, however, 
no longer a straight line but a curvy one.
It starts from $\xi_1$
and ends at
\begin{eqnarray}
\xi_2 &=& F^{\kappa}_{\sigma_1}
(\sqrt{\kappa} \sigma_1 + \xi_2^{(1)})
\nonumber\\
&=& F^{\kappa}_{\sigma_1}
(\sqrt{\kappa} \sigma_1
+F^{\kappa}_{\sigma_2}(\sqrt{\kappa} \sigma_2)).
\label{eqn:gamma2}
\end{eqnarray}
We have then a set of two points
$(\xi_1, \xi_2)$ in $\H$.
Set $n \geq 1$ and now we assume that
a realization of RW on $\Z$ up to time 
$n$ is specified by the series of $n$ steps,
$\vsigma(n)=(\sigma_1, \sigma_2, \dots, \sigma_n)$.
Let
\begin{equation}
S^{\kappa}_{\sigma}(z)=F^{\kappa}_{\sigma}(\sqrt{\kappa} \sigma+z)
\label{eqn:S1}
\end{equation}
for $\kappa > 0, \sigma \in \{-1,1\}$ and $z \in \H$.
We perform the following iteration of SCTs,
\begin{eqnarray}
{\cal S}^{\kappa}_{\vsigma(n)}(z)
&=& S^{\kappa}_{\sigma_1} \circ
S^{\kappa}_{\sigma_{2}} \circ \cdots
\circ S^{\kappa}_{\sigma_n}(z)
\nonumber\\
&\equiv&
S^{\kappa}_{\sigma_1}\Big(S^{\kappa}_{\sigma_{2}}
( \cdots ( S^{\kappa}_{\sigma_n}(z)) \cdots ) \Big).
\label{eqn:S2}
\end{eqnarray}
Then we have a curve consisting of a straight slit 
between the origin $\xi_0 = 0$ and $\xi_1$
and $n-1$ segments of curvy slits, which are sequentially
connected at $\xi_j, 1 \leq j \leq n-1$,
and the tip is at $\xi_n$.
See Fig. \ref{fig:fig_ISCT}.
In the present paper we call
${\cal S}^{\kappa}_{\vsigma(n)}(z)$
the iterative Schwarz-Christoffel transformation
(ISCT for short) driven by RW
specified by $\vsigma(n)=(\sigma_1, \dots, \sigma_n)$.

For $n=1,2,3, \dots$, we define the points in $\H$ by
\begin{equation}
\xi_n = {\cal S}^{\kappa}_{\vsigma(n)}(0).
\label{eqn:gammat}
\end{equation}
We set $\xi_0 \equiv 0$.
The sequence of points
$(\xi_0, \xi_1, \xi_2, \dots, \xi_n)$
is interpolated by straight lines.
We call it an ISCT path in $\H$
and denote it by $\Xi(0,n]$.
In other words, each realization $\vsigma(n)$ of RW
on $\Z$ is mapped to a path $\Xi(0,n]$
on $\H$ by the ISCT.

For a given $n \geq 1$, we introduce the following
recurrence relations for a series
$(\xi_n^{(0)}, \xi_n^{(1)}, \dots, \xi_n^{(n)})$,
\begin{equation}
\xi_n^{(k+1)}=S^{\kappa}_{\sigma_{n-k}}(\xi^{(k)}_{n}),
\quad k=0, 1,2, \dots, n-1
\label{eqn:git1}
\end{equation}
with $\xi_n^{(0)}=0$. 
Since $\sigma_j, j=1,2,3, \dots, n$ are i.i.d.,
the recurrence formula (\ref{eqn:git1})
is useful to calculate the position $\xi_n$,
which is given by $\xi_n^{(n)}$.
Moreover, if we introduce a parameter
$\theta \in (0, \pi/2)$ with the relation
\begin{equation}
\cos \theta=\sqrt{
\frac{\kappa}{\kappa+16}},
\label{eqn:theta1}
\end{equation}
(\ref{eqn:alpha1}) is written as
\begin{eqnarray}
\alpha^{\kappa}(\sigma)
&=& \frac{1}{2}(1-\sigma \cos \theta)
\nonumber\\
&=& \left\{ \begin{array}{ll}
\sin^2(\theta/2),
& \quad \sigma=1 \cr
\cos^2(\theta/2), 
& \quad \sigma=-1,
\end{array} \right.
\label{eqn:alpha2}
\end{eqnarray}
and (\ref{eqn:gamma1}) is given by
\begin{eqnarray}
\xi_1 &=& S^{\kappa}_{\sigma_1}(0)
\nonumber\\
&=& \left\{ \begin{array}{ll}
2 (\cot(\theta/2))^{\cos 2 \theta}
e^{i \pi \sin^2(\theta/2)},
& \quad \sigma_1=1 
\cr
2 (\cot(\theta/2))^{\cos 2 \theta}
e^{i \pi \cos^2(\theta/2)},
& \quad \sigma_1=-1.
\end{array} \right.
\label{eqn:gamma1b}
\end{eqnarray}
Note that $\xi_1(\sigma=-1)
=- (\xi_1(\sigma=1))^{*}$,
where $^{*}$ indicates complex conjugate.
The recurrence relation (\ref{eqn:git1}) is then
written as
\begin{equation}
\xi_n^{(k+1)} =
\left(\xi_{n}^{(k)}+\sigma_{n-k} \cot \frac{\theta}{2}
\right)^{\cos^2(\theta/2)}
\left( \xi_{n}^{(k)}-\sigma_{n-k} \tan \frac{\theta}{2}
\right)^{\sin^2(\theta/2)},
\quad k=0, 1,2, \dots, n-1
\label{eqn:git2}
\end{equation}
for $\sigma_{n-k}=\pm 1$ with $\xi_n^{(0)}=0$.
The expressions for the relations
between two real components of complex variable
$\xi_{n}^{(k+1)}$ and those of $\xi_{n}^{(k)}$ are
given in Appendix A.

\section{Networks on $\H$ generated by RW paths}

\begin{figure}[htbp]
  \begin{center}
   \includegraphics[width=100mm]{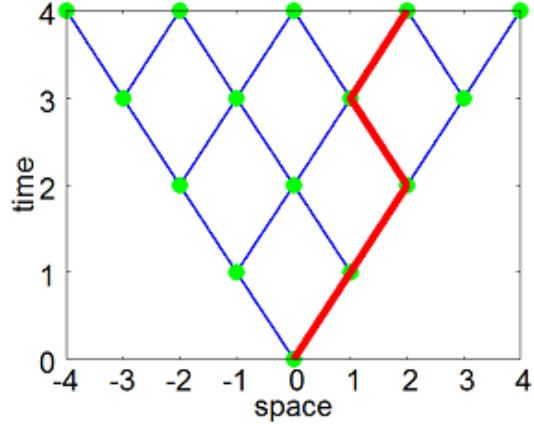}
  \end{center}
  \caption{Lattice $\Lambda_4$ and
  one realization of RW with
  $\vsigma(4)=(1,1,-1,1)$.}
  \label{fig:fig_RW4}
\end{figure}

For a fixed time period $n >0$,
consider a collection of all realization of steps of RW
on $\Z$,
\begin{equation}
{\cal F}(n)=\Big\{
\vsigma(n)=(\sigma_1, \sigma_2, \dots, \sigma_n):
\sigma_j \in \{-1,1\}, 1 \leq j \leq n \Big\}.
\label{eqn:cF1}
\end{equation}
The total number of realizations of RWs is
$|{\cal F}(n)|=2^n$.
We note that each realization of RW up to time $n >0$
is represented by
a directed path on a squared lattice 
in a triangular region in the spatio-temporal plane,
\begin{equation}
\Lambda_n= \Big\{ (j,k) \in \Z \times \{0,1,2, \dots \} :
0 \leq k \leq n, -k \leq j \leq k,
j+k=\mbox{even} \Big\}.
\label{eqn:Lambda1}
\end{equation}
Here any edge connecting nearest neighbor vertices
in $\Lambda_n$ is assumed to be directed in the positive direction
of the time axis; $(j,k) \to (j\pm 1, k + 1)$,
and each path is a sequence of edges all in the
positive direction, starting from
$(0,0)$ to $(j, n)$
with $-n \leq j \leq n$.
Figure \ref{fig:fig_RW4} shows $\Lambda_4$ and an example 
of a realization of RW with
$\vsigma(4)=(1,1,-1,1)$.

\begin{figure}[htbp]
  \begin{center}
   \includegraphics[width=80mm]{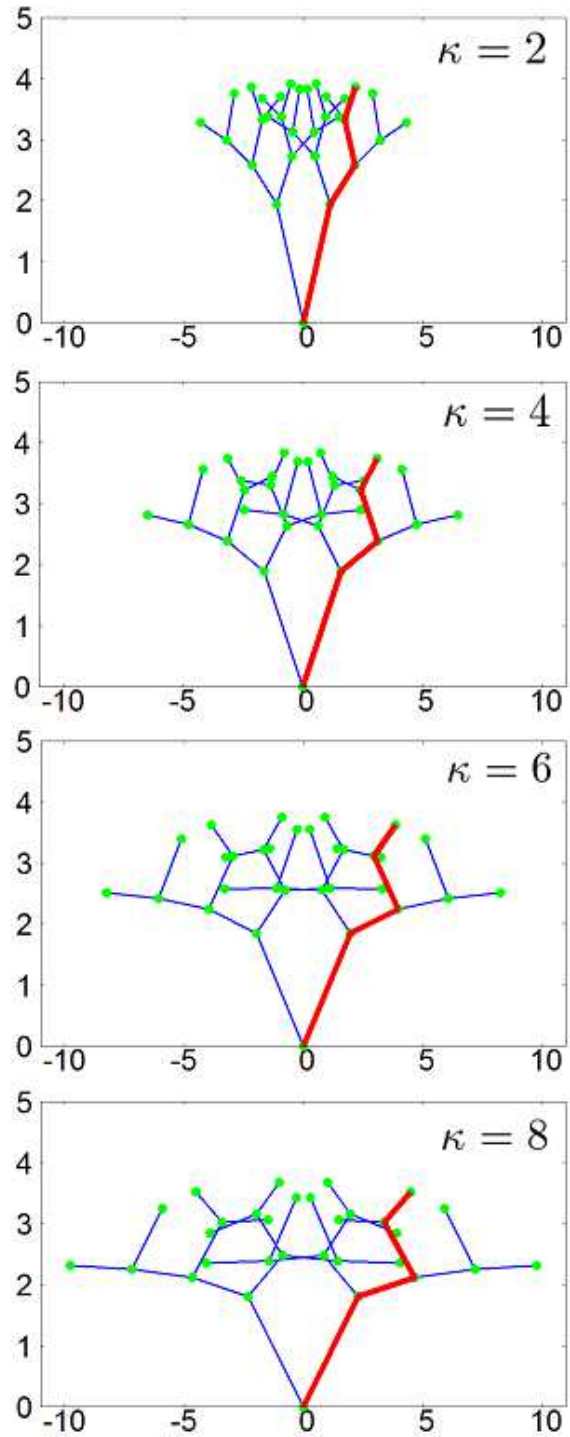}
  \end{center}
  \caption
  {Networks ${\cal N}_n^{\kappa}$ on $\H$
  for $n=4$ with $\kappa=2,4,6,8$.
  The ISCT paths corresponding to
  the same realization of RW
  $\vsigma(4)=(1,1,-1,1)$ are indicated by bold lines.}
  \label{fig:fig_Xi4}
\end{figure} 

\begin{figure}[htbp]
  \begin{center}
   \includegraphics[width=100mm]{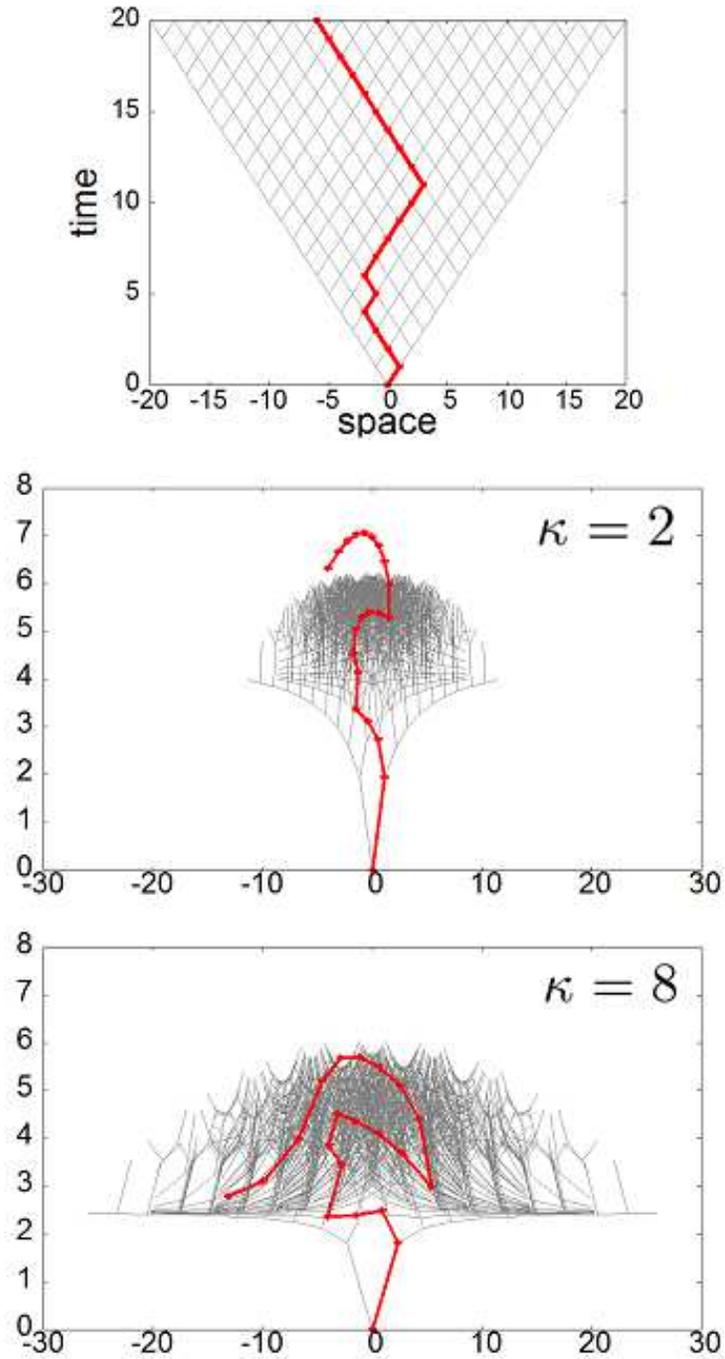}
  \end{center}
  \caption
  {(a) A realization of RW with time step
  $n=20$. (b) The ISCT path for $\kappa=2$
  generated from the RW shown in (a)
  with the network ${\cal N}_{10}^{2}$.
  (c) The ISCT path for $\kappa=8$
  generated from the same RW shown in (a)
  with the network ${\cal N}_{10}^{8}$.
  }
  \label{fig:fig_Xi20}
\end{figure} 

By collecting all ISCT paths up to time $n$, $\Xi(0,n]$,
we have a network on $\H$,
\begin{equation}
{\cal N}_n^{\kappa}
=\Big\{\Xi(0, n] :
\Xi_j=S^{\kappa}_{\vsigma(j)}(0), 1 \leq j \leq n,
\vsigma(n) \in {\cal F}(n) \Big\}.
\label{eqn:Gamma1}
\end{equation}
Figure \ref{fig:fig_Xi4} shows ${\cal N}_{n}^{\kappa}$ for
$\kappa=2,4,6$ and 8 for $n=4$.
As the parameter $\kappa$ increases,
the network becomes spreading wider in $\H$.
There the ISCT path corresponding to
the realization of RW
$\vsigma(4)=(1,1,-1,1)$ shown in 
Fig. \ref{fig:fig_RW4} is indicated
by a bold line for each value of $\kappa$.
In Fig. \ref{fig:fig_Xi20}, we compare the ISCT paths
with (b) $\kappa=2$ and (c) $\kappa=8$
both obtained from the same realization (a)
of RW with $n=20$ steps,
where the networks up to $n=10$,
${\cal N}_{10}^{\kappa}$,
are also shown in the background for each $\kappa$.
We can see that the ISCT path with $\kappa=8$ is
much more complicated than the path with $\kappa=2$.
As shown by Fig.\ref{fig:fig_Xi4} and Fig.\ref{fig:fig_Xi20},
the network ${\cal N}_n^{\kappa}$ is
bounded by the rightmost path $\Xi^{\max}(0, n]$
generated by $\vsigma_{\max}(n)=(1,1, \dots, 1)$
and the leftmost path  $\Xi^{\min}(0, n]$
generated by $\vsigma_{\min}(n)=(-1,-1, \dots, -1)$.
The height of $\Xi^{\max}_n$ and $\Xi^{\min}_n$,
$H_n^{\kappa}={\rm Im} \, \Xi^{\max}_n={\rm Im} \, \Xi^{\min}_n$,
is observed to converge to a positive constant
$H^{\kappa}_{\infty}$ in $n \to \infty$.
The numerical values are given by
$H^{\kappa}_{\infty}=0.014 (\kappa=1)$,
0.010 ($\kappa=2$),
0.0084 ($\kappa=3$),
0.0074 ($\kappa=4$),
0.0067 ($\kappa=5$),
0.0062 ($\kappa=6$),
0.0058 ($\kappa=7$),
and
0.0055 ($\kappa=8$).

\section{Numerical Analysis of curves
generated by ISCT}

For a given number of steps $n=NT$ of RW, we have defined
the rescaled ISCT path $\zeta^N(0,T]$ by
Eq.(\ref{eqn:zetaN}).
In particular, $\zeta(0,1]$ is obtained by
interpolating the series of points
$\{\xi_j/\sqrt{N}\}_{j=0}^N$ by straight lines.
We have studied statistical properties of
the rescaled ISCT paths based on
the numerical data of large scaled
computer simulations.

\subsection{Fractal Dimensions}

\begin{figure}[htbp]
  \begin{center}
   \includegraphics[width=100mm]{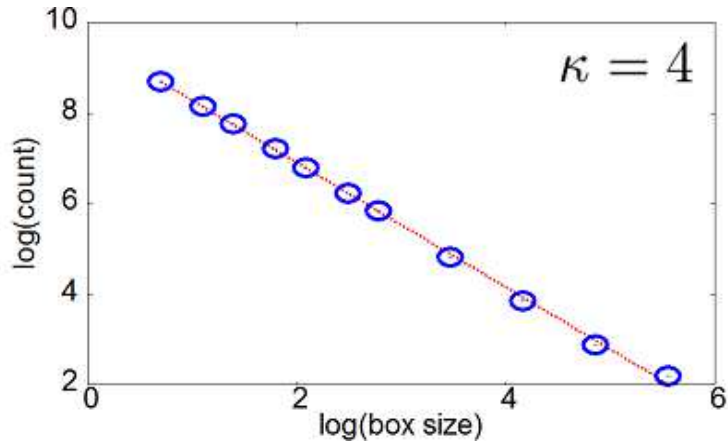}
  \end{center}
  \caption{Log-log plot of the box-counting data
  for the restricted ISCT paths for $\kappa=4$.}
  \label{fig:fig_boxcount_k4}
\end{figure}

Figure \ref{fig:fig_boxcount_k4} shows
a log-log plot of the box counting of segments
of $\zeta^N(0,1]$ with respect to the
box sizes for $\kappa=4$ with $N=5 \times 10^{4}$.
As shown by this figure, 
the data for any $\kappa$ can be fitted by
a straight line very well
and we can evaluate the approximate values of
fractal dimensions for finite $N$.

\begin{figure}[htbp]
  \begin{center}
   \includegraphics[width=100mm]{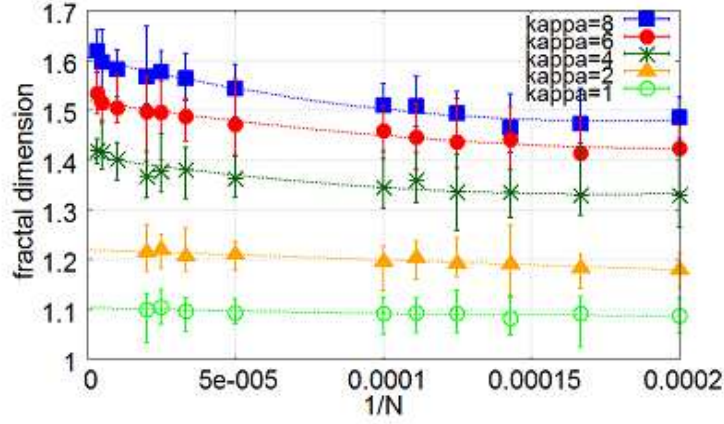}
  \end{center}
  \caption{Extrapolation of fractal dimensions
  in the $N \to \infty$ limit.}
  \label{fig:fig_dim1}
\end{figure}

The evaluated values 
up to at most $N = 3 \times 10^5$ 
are then plotted versus
$1/N$ in Fig.\ref{fig:fig_dim1}.
For each evaluation we prepared
$M=20$ samples, and the ranges of scattering of results
are shown by error bars in the figure.
There $N \to \infty$ limits
are extrapolated by three-parameter
fittings ; $d=a_0+a_1/N+a_2/N^2$.
The obtained values $a_0$ by the extrapolation
are denoted by $d^{\kappa}_{\rm ISCT}$.

\begin{figure}[htbp]
  \begin{center}
   \includegraphics[width=100mm]{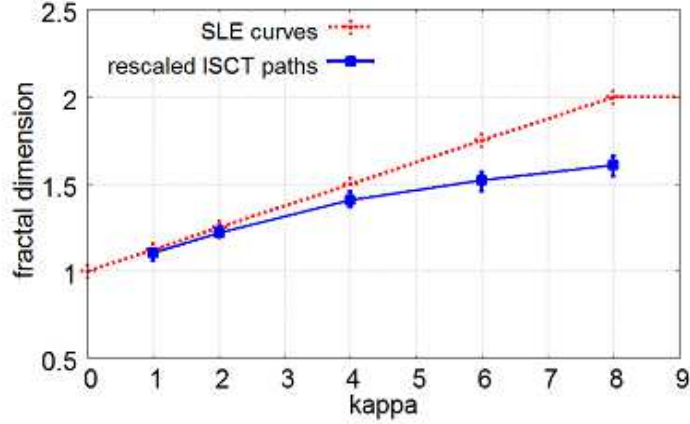}
  \end{center}
  \caption{Dependence of the fractal dimensions
  $d^{\kappa}_{\rm ISCT}$ of rescaled ISCT paths
  is shown.
  The Hausdorff dimensions of the SLE$_{\kappa}$ curves
  given by Eq.(\ref{eqn:d1}) are also plotted
  by a dotted line.}
  \label{fig:fig_dim2}
\end{figure}

Figure \ref{fig:fig_dim2} shows dependence 
of $d^{\kappa}_{\rm ISCT}$ on $\kappa$. 
The Hausdorff dimensions of the SLE$_{\kappa}$
curves given by (\ref{eqn:d1}) are also shown
by a dotted line.
Systematic deviation is found
between $d^{\kappa}$ for the SLE curves
and $d^{\kappa}_{\rm ISCT}$ numerically evaluated
for the rescaled ISCT paths.
We observe in Fig.\ref{fig:fig_dim1} that
the approximate value of fractal dimension
for finite $N$ is increasing 
as $N$ is increasing,
and the ratio of increment becomes larger
as $\kappa$ approaches the value 8.
So we expect that the deviation will be
systematically reduced,
if we can perform numerical simulation 
for larger $N$'s and make appropriate 
extrapolation to the $N \to \infty$ limit.

\subsection{Generalized Cardy's Formula}

\begin{figure}[htbp]
  \begin{center}
   \includegraphics[width=100mm]{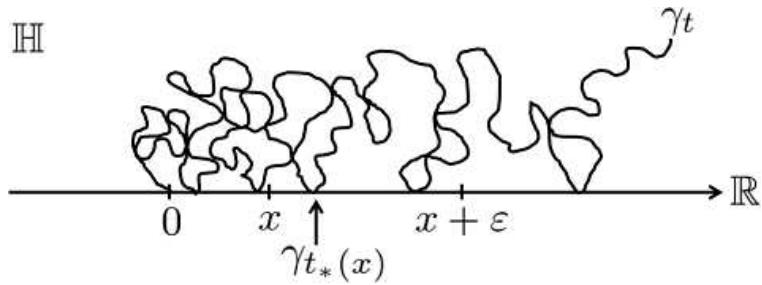}
  \end{center}
  \caption{Illustration of the point $\gamma_{t_*(x)}$
  on the real axis $\R$ for an SLE$_{\kappa}$
  curve with $4 < \kappa < 8$.}
  \label{fig:fig_Cardy1}
\end{figure}

Here we first consider the SLE curve 
$\gamma(0, \infty)=\{\gamma_t : 0 < t < \infty \}$ in the case
\begin{equation}
4 < \kappa < 8.
\label{eqn:kappa1}
\end{equation}
In this case, $\gamma_t$ starting from the origin
will hit the real axis $\R$ infinitely many times
\cite{Law05,Law07}.
For $x > 0$ we can define
\begin{eqnarray}
t_{*}(x) &=& \mbox{the first time, when $\gamma_t$ hits
a point in $[x, \infty)$ on $\R$}.
\label{eqn:tstar}
\end{eqnarray}
Then, 
$\gamma_{t_{*}(x)}$ is the leftmost point
in the interval $[x, \infty)$,
at which $\gamma_t$ hits $\R$.
Note that, if $\kappa \leq 4$, $t_{*}(x)=\infty$,
and if $\kappa \geq 8$, $\gamma_{t_*(x)}=x$, 
with probability one.
For (\ref{eqn:kappa1}), $\gamma_{t_*(x)}$
has a nontrivial distribution.
For each $\varepsilon > 0$, we observe whether
$\xi_{t_*(x)} < x+ \varepsilon$ or
$\xi_{t_*(x)} \geq x+\varepsilon$.
Figure \ref{fig:fig_Cardy1} illustrates the former case.
For the SLE$_{\kappa}$ curves with (\ref{eqn:kappa1}), 
the following formula is established.
(See Proposition 6.34 in \cite{Law05}.)
\begin{eqnarray}
&& {\rm P}(\gamma_{t_*(x)} < x+\varepsilon )
= \frac{\Gamma(4/\kappa)}{\Gamma(8/\kappa-1)\Gamma(1-4/\kappa)}
\int_{0}^{\varepsilon/(x+\varepsilon)} u^{8/\kappa-2}
(1-u)^{-4/\kappa} du
\nonumber\\
\label{eqn:Cardy1}
&& \qquad = \frac{\Gamma(4/\kappa)}
{\Gamma(8/\kappa) \Gamma(1-4/\kappa)}
\left(\frac{\varepsilon}{x+\varepsilon} \right)^{8/\kappa-1}
F\left( \frac{4}{\kappa}, \frac{8}{\kappa}-1,
\frac{8}{\kappa}; \frac{\varepsilon}{x+\varepsilon} \right),
\end{eqnarray}
where $\Gamma(z)$ is the gamma function and
$F(\alpha, \beta, \gamma; z)$ is Gauss's 
hypergeometric function.
This formula can be regarded as a generalization of
Cardy's formula, since the original formula
corresponding to (\ref{eqn:Cardy1}) with $\kappa=6$
was derived by Cardy \cite{Car92} for 
the ``percolation exploration process" in the
critical percolation model, 
and then the continuum limit of that process was proved to 
be described by
the SLE curve with $\kappa=6$ \cite{Smir01,Bef07}.
For $0 < \varepsilon \ll 1$, the above formula
gives a power law
\begin{equation}
 {\rm P}(\gamma_{t_*(x)} < x+\varepsilon )
 \simeq \left(\frac{\varepsilon}{x}\right)^{\delta(\kappa)}
\label{eqn:Cardy2}
\end{equation}
with the exponent
\begin{equation}
\delta(\kappa)=\frac{8-\kappa}{\kappa}
\label{eqn:Cardy3}
\end{equation}

Now we consider the ISCT paths.
For integers $N$ and $T$ with $N \gg 1$,
prepare a realization of RW represented by
$\vsigma(NT)=(\sigma_1, \sigma_2, \dots, \sigma_{NT})$.
For each $n=1,2, \dots, NT$, by using the data
$(\sigma_{1}, \sigma_{2}, \dots, \sigma_{n})$,
we calculate the position $\xi_n$ in $\H$
following the recurrence formula (\ref{eqn:git1}).
As noted at the end of Sec.III, 
${\rm Im} \, \xi_n >0$ for any $n \geq 1$.
So we set a small value $h >0$ and look for the event
\begin{equation}
{\cal E}^{N}_{h,x}(n)=\left\{
\frac{{\rm Im} \, \xi_n}{\sqrt{N}} < h
\quad \mbox{and} \quad
\frac{{\rm Re} \, \xi_n}{\sqrt{N}} \geq x \right\}.
\label{eqn:event}
\end{equation}
We define
\begin{equation}
n_h(x)=\min \Big\{n: 1 \leq n \leq NT,
\, \mbox{${\cal E}^{N}_{h, x}(n)$ occurs} \Big\}.
\label{eqn:jh}
\end{equation}
If $n_h(x) \leq NT$, we define
$t_h(x)=n_h(x)/N$ and calculate the value
${\rm Re} \, \zeta^{N}_{t_h(x)}
={\rm Re} \, \Xi^{N}_{N t_h(x)}/\sqrt{N}
={\rm Re} \, \xi^N_{n_h(x)}/\sqrt{N}$.
If $n_{h}(x) > NT$, that is,
the event (\ref{eqn:event}) does not occur for 
the given $\vsigma(NT)$, then $t_h(x) > T$.
The probability distribution function
for the rescaled ISCT paths,
which corresponds to
$P(\gamma_{t_*(x)} < x+\varepsilon)$, may be given by
\begin{equation}
\lim_{h \to 0} \lim_{T \to \infty} \lim_{N \to \infty}
{\rm P} \Big( {\rm Re} \, \zeta^N_{t_h(x)} < x + \varepsilon,
\, t_h(x) \leq T \Big)
\equiv {\rm P}(\zeta_{t_*(x)} < x + \varepsilon),
\label{eqn:Prob1}
\end{equation}
where $\zeta(0, \infty)
=\lim_{T \to \infty} \lim_{N \to \infty} \zeta^N(0,T]$.

\begin{figure}[htbp]
  \begin{center}
   \includegraphics[width=100mm]{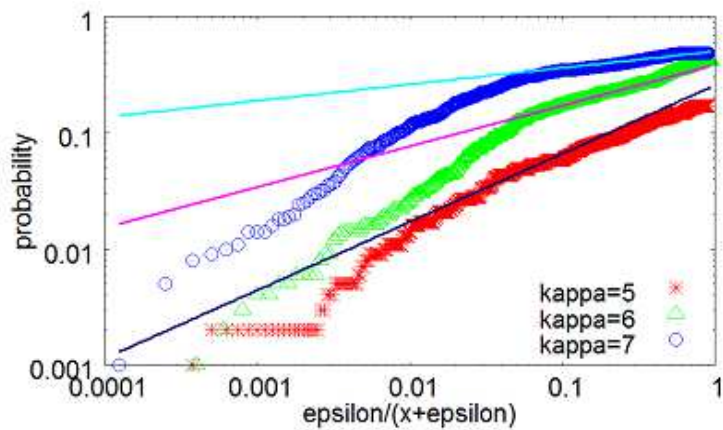}
  \end{center}
  \caption{Log-log plots of the numerical evaluations
  of the probability 
  ${\rm P}({\rm Re} \, \zeta^{N}_{t_h(x)} < x + \varepsilon,
  \, t_h(x) < 1)$
  versus $\varepsilon/(x+\varepsilon)$ for the rescaled ISCT paths.}
  \label{fig:fig_Cardy2}
\end{figure}

\begin{figure}[htbp]
  \begin{center}
   \includegraphics[width=100mm]{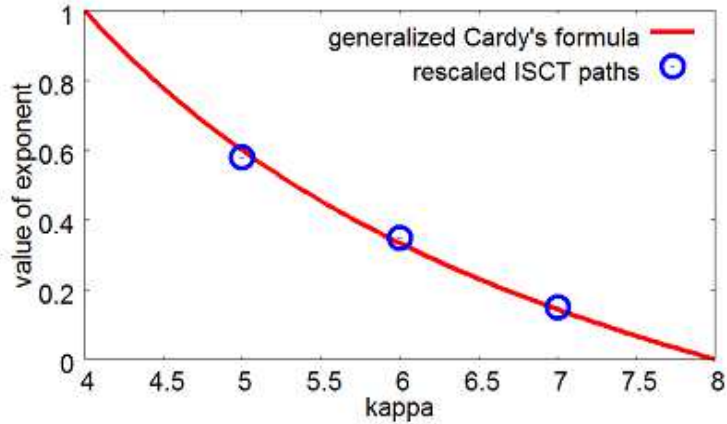}
  \end{center}
  \caption{Numerical evaluations of the exponent $\delta$
  in the power law (\ref{eqn:power2}) for $\kappa=5,6$, and 7.
  The curve shows Eq.(\ref{eqn:Cardy3})
  of the generalized Cardy's formula.}
  \label{fig:fig_Cardy3}
\end{figure}

In numerical calculations, we have set
$N=10^4$ and $T=1$
and prepared $M=1000$ realizations of RW, which are
randomly generated.
Since $T=1$, we are allowed to consider only small
values of $x$ and $\varepsilon$.
The threshold value $h$ is wanted to be small,
but for finiteness of $N$, it should be positive.
In Fig.\ref{fig:fig_Cardy2}, we show log-log plots
of the numerical evaluations of
${\rm P}({\rm Re} \, \zeta^{N}_{t_h(x)} < x + \varepsilon, \,
t_h(x) \leq 1)$
for $\kappa=5$ with $x=0.8, h=0.1$,
$\kappa=6$ with $x=0.5, h=0.1$,
and
$\kappa=7$ with $x=0.8, h=0.2$.
For finiteness of $N$ and smallness of the number of 
samples $M$, data scatter for small $\varepsilon$.
We find, however, power-law behaviors in the intermediate
regions of $\varepsilon$;
\begin{equation}
 {\rm P}(\zeta_{t_*(x)} < x+\varepsilon )
 \simeq \varepsilon^{\delta}.
\label{eqn:power2}
\end{equation}
By linear fitting as shown in Fig.\ref{fig:fig_Cardy2},
we have evaluated the values of exponent $\delta$.
The results are plotted in Fig.\ref{fig:fig_Cardy3},
where Eq.(\ref{eqn:Cardy3}) derived from
the generalized Cardy's formula
is also shown by a curve.
The good agreement implies that
in the proper limit (\ref{eqn:Prob1})
will also follow the generalized Cardy's formula
in the parameter region (\ref{eqn:kappa1}).

\section{Concluding Remarks}

In the present paper, we have proposed an algorithm,
which generates a random discrete path
$\Xi(0,NT]$ on the upper half complex-plane $\H$
as a functional of a path of random walk
$W(0,NT]$ on the real axis $\R$ for integers 
$N$ and $T$.
The system has one parameter $\kappa >0$
and we call the path an ISCT$_{\kappa}$ path.
We have studied the rescaled ISCT$_{\kappa}$ path
defined by
$\zeta^N(0,T]=\Xi(0,NT]/\sqrt{N}$
for large $N$ by computer simulations.
The numerical analysis of the distributions
of $\zeta^{N}(0,T]$ supports our expectation
that the limits of the rescaled ISCT$_{\kappa}$ paths
$\zeta(0, \infty)=\lim_{T \to \infty}
\lim_{N \to \infty} \zeta^N(0,T]$
will have the same statistical properties
as the SLE$_{\kappa}$ curves have.

The rescaled ISCT$_{\kappa}$ paths
$\zeta^N(0,T]$ can be regarded as 
discrete approximations of the SLE$_{\kappa}$ 
curves.
In this sense, the present study could be
included by the previous numerical work
\cite{Ken07,Ken09}.
In this paper, however, we have emphasized on 
our interest in the ISCT itself
as a simple algorithm to generate
complicated discrete dynamics of a point
on $\H$.
Dependence on the parameter $\kappa$
of complexity of the SLE curves
is demonstrated
by dependence of complexity of the
network ${\cal N}_{n}^{\kappa}$
of the ISCT paths
on the angle $\alpha \pi$ of the slit
generated by a single SCT.

We have learned that stochastic analysis
is necessary and useful to study 
statistics and stochastics 
of the SLE$_{\kappa}$ curves \cite{Law05,Law07}.
Although we have reported only numerical study
in this paper, we hope that
the combinatorics and
statistical mechanics methods developed
for solvable models on lattices
will be useful to analyze statistics and stochastics
of the ISCT$_{\kappa}$ paths on $\H$,
since they are functionals of simple
random walks in one dimension.

\begin{acknowledgments}
The present authors would like to thank
M. Matsushita and N. Kobayashi for useful
discussion on application of the
fractal analysis to the present work.
This work was partially supported by the Grant-in-Aid
for Scientific Research (C) (No.21540397) of
Japan Society for the Promotion of Science.
\end{acknowledgments}

\appendix
\section{Recurrence Relations}

Let $\xi_{n}^{(k)}=2 r e^{i \phi},
r >0, 0 < \phi < \pi$ for $k < n$.
Then (\ref{eqn:git2}) gives
\begin{equation}
\xi_n^{(k+1)}=2 R_{\sigma_{n-k}} \exp(i \Phi_{\sigma_{n-k}}),
\quad \sigma_{n-k}= \pm 1
\label{eqn:gammajA1}
\end{equation}
with
\begin{eqnarray}
R_{\sigma} &=&
2\left(r^2+2 \sigma r \cos{\phi} \cot{\frac{\theta}{2}}
+\cot^2{\frac{\theta}{2}}\right)
^{(\cos^2(\theta/2))/2} \nonumber\\
\label{eqn:R1}
&& \times
\left(r^2-2 \sigma r \cos{\phi} \tan{\frac{\theta}{2}}
+\tan^2 \frac{\theta}{2} \right)
^{(\sin^2(\theta/2))/2},
\\
\Phi_{\sigma} &=&
\arccos\left({\frac{r\cos{\phi}
+\sigma \cot(\theta/2)}
{\sqrt{r^2+2 \sigma r \cos{\phi} \cot(\theta/2)+\cot^2 (\theta/2)}}}
\right){\cos^2{\frac{\theta}{2}}}
\nonumber\\
\label{eqn:Phi1}
&& +
\arccos\left({\frac{r\cos{\phi}-\sigma \tan{(\theta/2)}}
{\sqrt{r^2-2 \sigma r \cos{\phi}
\tan{(\theta/2)} +\tan^2(\theta/2)}}}
\right){\sin^2{\frac{\theta}{2}}}
\end{eqnarray}

When we set $\xi_{n}^{(k)}=2(x+iy),
x \in \R, y >0, k < n$,
the above gives
\begin{equation}
\xi_n^{(k+1)}=2(X_{\sigma_{n-k}}+i Y_{\sigma_{n-k}}),
\quad \sigma_{n-k}=\pm 1
\label{eqn:gammajA2}
\end{equation}
with
\begin{eqnarray}
X_{\sigma} &=& 
\left(x^2+y^2+2 \sigma x \cot{\frac{\theta}{2}}
+\cot^2{\frac{\theta}{2}}\right)
^{(\cos^2(\theta/2))/2}
\left(x^2+y^2-2 \sigma x \tan{\frac{\theta}{2}}
+\tan^2{\frac{\theta}{2}}\right)
^{(\sin^2(\theta/2))/2}
\nonumber\\
&&\quad \times \cos \Biggl[
\arccos\left({\frac{x+\sigma \cot{(\theta/2)}}
{\sqrt{x^2+y^2+2 \sigma x \cot{(\theta/2)}
+\cot^2{(\theta/2)}}}}
\right){\cos^2{\frac{\theta}{2}}}
\nonumber\\
\label{eqn:X1}
& & \qquad \qquad 
+\arccos\left({\frac{x-\sigma \tan{(\theta/2)}}
{\sqrt{x^2+y^2-2 \sigma x \tan{(\theta/2)}
+\tan^2{(\theta/2)}}}}\right){\sin^2{\frac{\theta}{2}}}
\Biggr],
\\
Y_{\sigma} &=&
\left(x^2+y^2+2 \sigma x \cot{\frac{\theta}{2}}
+\cot^2{\frac{\theta}{2}}\right)
^{(\cos^2(\theta/2))/2}
\left(x^2+y^2-2 \sigma x \tan{\frac{\theta}{2}}
+\tan^2{\frac{\theta}{2}}\right)
^{(\sin^2 (\theta/2))/2}
\nonumber\\
&& \quad \times \sin \Biggl[
\arccos\left({\frac{x+\sigma \cot{(\theta/2)}}
{\sqrt{x^2+y^2+2 \sigma x \cot{(\theta/2)}
+\cot^2{(\theta/2)}}}}
\right){\cos^2{\frac{\theta}{2}}}
\nonumber\\
\label{eqn:Y1}
&& \qquad \qquad 
+\arccos\left({\frac{x-\sigma \tan{(\theta/2)}}
{\sqrt{x^2+y^2 -2 \sigma x \tan{(\theta/2)}
+\tan^2 {(\theta/2)}}}}
\right){\sin^2{\frac{\theta}{2}}}
\Biggr].
\end{eqnarray}


\begin{thebibliography}{99}
\bibitem{SLEb}
Usually the time evolution of the inverse map
$g_t=f_t^{-1}$ given by
$\partial g_t/\partial t=2/(g_t-\sqrt{\kappa} B_t)$
is called the SLE.
Since $f_t$ is one-to-one,
$\partial f_t/\partial z \not=0$,
$f_t$ can determine $g_t$ completely.
\end{thebibliography}

\end{document}